\documentclass[aps,prd,showpacs]{revtex4}

\usepackage{amsmath,amsfonts,amssymb}

\allowdisplaybreaks[1]

\newcommand{\ie}{{\it i.e.}}

\newcommand{\eq}{Eq.}
\newcommand{\eqs}{Eqs.}

\newcommand{\Ref}{Ref.}
\newcommand{\Refs}{Refs.}
\newcommand{\Sec}{Sec.}

\begin{document}

\title{Bimaximal fermion mixing from the quark and leptonic mixing matrices}

\author{Tommy Ohlsson}\email{tommy@theophys.kth.se}
\affiliation{Division of Mathematical Physics, Department of Physics,
School of Engineering Sciences, Royal Institute of Technology (KTH) --
AlbaNova University Center, Roslagstullsbacken 11, 106 91 Stockholm,
Sweden}

\begin{abstract}
In this paper, we show how the mixing angles of the standard
parameterization add when multiplying the quark and leptonic mixing
matrices, \ie, we derive explicit sum rules for the quark and leptonic
mixing angles. In this connection, we also discuss other recently
proposed sum rules for the mixing angles assuming bimaximal fermion
mixing. In addition, we find that the present experimental and
phenomenological data of the mixing angles naturally fulfill our sum
rules, and thus, give rise to bilarge or bimaximal fermion mixing.
\end{abstract}


\maketitle

\vspace{-5mm}
\section{Introduction}

One of the most fundamental questions in particle physics that still
remains to be answered is ``What is the mixing of quarks and
leptons?''. Here this question will be addressed in a phenomenological
way. The mixings of both quarks and leptons are described by unitary
mixing matrices (relating the fields in the flavor and mass
bases). Assuming three generations of quarks and leptons, the quark
mixing matrix is known as the Cabibbo--Kobayashi--Maskawa (CKM) matrix
\cite{Cabibbo:1963yz,Kobayashi:1973fv}, whereas the leptonic mixing
matrix is mostly known as the Maki--Nakagawa--Sakata (MNS) matrix
\cite{Maki:1962mu}. Using the so-called standard parameterization of
these two matrices \cite{Eidelman:2004wy}, it turns out that the CKM
matrix has three small mixing angles ($\theta_{12}^{\rm CKM} =
13.0^\circ \pm 0.1^\circ$, $\theta_{13}^{\rm CKM} = 0.2^\circ \pm
0.1^\circ$, and $\theta_{23}^{\rm CKM} = 2.4^\circ \pm 0.1^\circ$
\cite{Gilman:2004pt}) and the MNS matrix has two large mixing angles
as well as probably one small mixing angle ($\theta_{12}^{\rm MNS} =
{33.2^\circ}^{+4.9^\circ}_{-4.6^\circ}$, $\theta_{13}^{\rm MNS} = 0
\pm 12.5^\circ$, and $\theta_{23}^{\rm MNS} =
{45.0^\circ}^{+10.6^\circ}_{-9.4^\circ}$ \cite{Maltoni:2004ei}). In
addition to the mixing angles, the standard parameterization also
contains a CP-violating phase $\theta_{\rm CP}$~\footnote{Note that a
general $3 \times 3$ unitary mixing matrix can also contain two
Majorana CP-violating phases. However, these two phases are not
measurable using neutrino oscillations and, in general, both Dirac and
Majorana CP-violating phases will make the situation much more
complicated than without them.}. In the quark sector, CP violation has
been measured restricting the CP-violating phase to $\delta_{\rm
CP}^{\rm CKM} = 1.05 \pm 0.24$ \cite{Gilman:2004pt}, whereas in the
leptonic sector, until now, CP violation has not been measured leading
to a completely undetermined value of the corresponding CP-violating
phase. Recently, it has been suggested that the mixing angles that
parametrize the mixing matrices could fulfill the following relations
\cite{Raidal:2004iw,Li:2005yj}: $\theta_{12}^{\rm CKM} +
\theta_{12}^{\rm MNS} = \tfrac{\pi}{4}$, $\theta_{13}^{\rm CKM} \sim
\theta_{13}^{\rm MNS} = {\cal O}(\lambda^3)$, and $\theta_{23}^{\rm
CKM} + \theta_{23}^{\rm MNS} = \tfrac{\pi}{4}$, where $\lambda$ is the
Wolfenstein parameter \cite{Wolfenstein:1983yz}. It should be noted
that the first relation, which relates the Cabibbo angle and the solar
mixing angle, was proposed a long time ago
\cite{Petcov:1993rk}. Nevertheless, this relation has recently been
discussed in the literature
\cite{Smirnov:2004ju,Raidal:2004iw,Minakata:2004xt,Frampton:2004vw,Ferrandis:2004vp,Kang:2005as,Cheung:2005gq,Xing:2005ur,Antusch:2005ca,Mohapatra:2005md,Li:2005yj}
to a great extent and it is now referred to as the quark-lepton
complementarity (QLC) relation. Next, these three above relations
suggest that the bilarge leptonic mixing and the small quark mixing
could be related to some exact fundamental bimaximal mixing. Thus, a
flavor symmetry is needed in order to describe such a mixing.  It
should be pointed out that the values of the mixing parameters of
course depend on the parameterization used. A relation for the mixing
parameters in a specific representation of a mixing matrix is not
necessarily the same in another representation.

In this paper, we argue that bimaximal mixing naturally appears when
multiplying the quark and leptonic mixing matrices coming from recent
experimental and phenomenological data.

The paper is organized as follows: In \Sec~\ref{sec:sumrules}, we
derive explicit sum rules for the quark and leptonic mixing angles resulting in
total mixing angles for a general fermion mixing. Next, we show
that the present experimental and phenomenological values of the quark
and leptonic mixing angles naturally lead to bilarge or maybe even
bimaximal fermion mixing. Then, in \Sec~\ref{sec:discussion}, we
discuss the earlier obtained results on QLC as well as bilarge or
bimaximal fermion mixing. Finally, in \Sec~\ref{sec:S&C}, we summarize
our results and present our conclusions.

\section{Explicit sum rules for quark and leptonic mixing angles}
\label{sec:sumrules}

The standard parameterization of a $3 \times 3$ unitary mixing matrix
is given by \cite{Eidelman:2004wy}
\begin{align}
U &= {\rm e}^{{\rm i} \lambda_7 \theta_{23}} U_\delta {\rm e}^{{\rm i}
  \lambda_5 \theta_{13}} U_\delta^\dagger {\rm e}^{{\rm i} \lambda_2
  \theta_{12}} = O_{23} U_\delta O_{13} U^\dagger_\delta O_{12}
  \nonumber\\ &= \left(
  \begin{matrix} c_{12} c_{13} & s_{12} c_{13} & s_{13} {\rm e}^{-{\rm
  i} \delta_{\rm CP}} \\ -s_{12} c_{23} - c_{12} s_{13} s_{23} {\rm
  e}^{{\rm i} \delta_{\rm CP}} & c_{12} c_{23} - s_{12} s_{13} s_{23}
  {\rm e}^{{\rm i} \delta_{\rm CP}} & c_{13} s_{23} \\ s_{12} s_{23} -
  c_{12} s_{13} c_{23} {\rm e}^{{\rm i} \delta_{\rm CP}} & -c_{12}
  s_{23} - s_{12} s_{13} c_{23} {\rm e}^{{\rm i} \delta_{\rm CP}} &
  c_{13} c_{23} \end{matrix} \right) \,,
   \label{eq:U}
\end{align}
where $\lambda_i$ ($i = 1,2,\ldots,8$) are the Gell-Mann matrices,
$O_{ij}$ is the orthogonal rotation matrix in the $ij$-plane which
depends on the mixing angle $\theta_{ij}$, $U_\delta=\mathrm{diag}
(1,1,{\rm e}^{{\rm i}\delta_{\rm CP}})$, $\delta_{\rm CP}$ being the
Dirac CP-violating phase, $s_{ij} \equiv \sin \theta_{ij}$, and
$c_{ij} \equiv \cos \theta_{ij}$. This parameterization is used for
both the quark and leptonic sectors. Now, we denote the
Cabibbo--Kobayashi--Maskawa quark mixing matrix by $V_{\rm CKM}$,
whereas we denote the Maki--Nakagawa--Sakata leptonic mixing matrix by
$U_{\rm MNS}$. In general, in the three-flavor case, there are also
two Majorana CP-violating phases, which can be introduced in a mixing
matrix $U$ by replacing the matrix with $U \Phi$, where $\Phi =
\mathrm{diag}(1,{\rm e}^{{\rm i}\phi_1},{\rm e}^{{\rm i}\phi_2})$ with
$\phi_1$ and $\phi_2$ being the Majorana CP-violating phases. However,
these two phases do not affect neutrino oscillations and will
therefore not be considered here.

The present experimental and phenomenological values of the moduli of
the quark and leptonic mixing matrices are given by
\cite{Gilman:2004pt,Maltoni:2004ei}
\begin{align}
|V_{\rm CKM}| & = \left( \begin{matrix} 0.9739 \div 0.9751 & 0.221
  \div 0.227 & 0.0029 \div 0.0045\\ 0.221 \div 0.227 & 0.9730 \div
  0.9744 & 0.039 \div 0.044\\ 0.0048 \div 0.014 & 0.037 \div 0.043 &
  0.9990 \div 0.9992 \end{matrix} \right), \label{eq:CKMvalues}\\
|U_{\rm MNS}| & = \left( \begin{matrix} 0.79 \div 0.88 & 0.48 \div
  0.62 & < 0.22 \\ 0.14 \div 0.64 & 0.36 \div 0.82 & 0.58 \div 0.84 \\
  0.14 \div 0.64 & 0.36 \div 0.82 & 0.58 \div 0.84 \end{matrix}
\right), \label{eq:MNSvalues}
\end{align}
where the values of the modulus of the quark mixing matrix are the
90~\% confidence level ranges, whereas the values of the modulus of
the leptonic mixing matrix are the 3$\sigma$ (99.7~\% confidence
level) ranges. In fact, the modulus of the leptonic mixing matrix has
been constructed from the following quantities $0.23 \leq \sin^2
\theta_{12}^{\rm MNS} \leq 0.38$, $\sin^2 \theta_{13}^{\rm MNS} \leq
0.047$, and $0.34 \leq \sin^2 \theta_{23}^{\rm MNS} \leq 0.68$
\cite{Maltoni:2004ei}. Another recent phenomenological value of this
matrix is given by \cite{Gonzalez-Garcia:2004jd}
\begin{equation}
|U_{\rm MNS}| = \left( \begin{matrix} 0.79 \div 0.88 & 0.47 \div
  0.61 & < 0.20 \\ 0.19 \div 0.52 & 0.42 \div 0.73 & 0.58 \div 0.82 \\
  0.20 \div 0.53 & 0.44 \div 0.74 & 0.56 \div 0.81 \end{matrix}
\right). \label{eq:MNSvaluesGG}
\end{equation}
Note that the ranges of the quark mixing matrix have been determined
using eight constraints from tree-level processes, which means that
there will be no information on the CP-violating phase in the quark
sector, and thus, the values of this phase can be set to zero, \ie,
$\delta_{\rm CP}^{\rm CKM} = 0$. Actually, in order to obtain
information on the CP-violating phase in this sector, we need to take
into account additional loop-level processes as additional constraints
\cite{Gilman:2004pt}. In addition, note that there is no knowledge
about the value of the CP-violating phase in the leptonic sector, \ie,
the value of $\delta_{\rm CP}^{\rm MNS}$ is allowed to lie in the
whole interval $[0,2\pi)$. {}From the two matrices $V_{\rm CKM}$ and
$U_{\rm MNS}$, assuming that the CP-violating phases in both the quark
and leptonic sectors are equal to zero, \ie, $\delta_{\rm CP}^{\rm
CKM} = 0$ and $\delta_{\rm CP}^{\rm MNS} = 0$, as well as using the
above ranges of the matrix elements in \eqs~(\ref{eq:CKMvalues}) and
(\ref{eq:MNSvalues}), we can read off the mixing angles to be
\cite{Ohlsson:2002rb}
$$
\begin{array}{ll}
\left\{
\begin{array}{l}
\theta_{12}^{\rm CKM} = 13.0^\circ \pm 0.1^\circ, \\
\theta_{13}^{\rm CKM} = 0.2^\circ \pm 0.1^\circ, \\
\theta_{23}^{\rm CKM} = 2.4^\circ \pm 0.1^\circ, \\
\end{array}
\right.
&
\left\{
\begin{array}{l}
\theta_{12}^{\rm MNS} = 33.2^\circ \pm 4.9^\circ, \\
\theta_{13}^{\rm MNS} = 0 \pm 12.5^\circ, \\
\theta_{23}^{\rm MNS} = 45.0^\circ \pm 10.6^\circ.\\
\end{array}
\right.
\end{array}
$$
Note that the matrix elements in \eq~(\ref{eq:MNSvaluesGG}) yield
the following values for the leptonic mixing angles: $\theta_{12}^{\rm
MNS} = 32.9^\circ \pm 4.8^\circ$, $\theta_{13}^{\rm MNS} = 0 \pm
11.5^\circ$, and $\theta_{23}^{\rm MNS} = 45.6^\circ \pm 10.1^\circ$,
which are more or less the same as the ones obtained above.

In order to investigate mixing on a more fundamental level, we will
add the quark and leptonic mixings. This is performed by multiplying
the corresponding mixing matrices, which is motivated by quark-lepton
unification. However, there are two possibilities of multiplying these
matrices either $U_{\rm MNS} V_{\rm CKM}$ or $V_{\rm CKM} U_{\rm
MNS}$, which have been investigated in \Ref~\cite{Li:2005yj}. Note
that these two resulting unitary mixing matrices do not commute, which
means that the two possible ways of multiplying the matrices will give
different results. Furthermore, note that the mixing angles do not
simply add in the trivial way as in the case of $2 \times 2$ unitary
(or orthogonal) mixing matrices, \ie, $\theta = \theta^{\rm MNS} +
\theta^{\rm CKM} = \theta^{\rm CKM} + \theta^{\rm MNS}$.

Multiplying the two unitary mixing matrices in the following order
\begin{equation}
W_1 = U_{\rm MNS} V_{\rm CKM}
\label{eq:W_1}
\end{equation}
and assuming that the quark mixing angles are small compared with the
leptonic mixing angles, we obtain series expansions for the total
mixing angles of the $W_1$ matrix
\begin{align}
\theta_{12} &\simeq \theta_{12}^{\rm MNS} + \theta_{12}^{\rm CKM} +
\left( s_{12}^{\rm MNS} \theta_{13}^{\rm CKM} - c_{12}^{\rm MNS}
\theta_{23}^{\rm CKM} \right) \tan \theta_{13}^{\rm MNS},
\label{eq:th12_1} \\
\theta_{13} &\simeq \theta_{13}^{\rm MNS} + c_{12}^{\rm MNS}
\theta_{13}^{\rm CKM} + s_{12}^{\rm MNS} \theta_{23}^{\rm CKM}, \\
\theta_{23} &\simeq \theta_{23}^{\rm MNS} + \left( c_{12}^{\rm MNS}
\theta_{23}^{\rm CKM} - s_{12}^{\rm MNS} \theta_{13}^{\rm CKM} \right)
\sec \theta_{13}^{\rm MNS}, \label{eq:th23_1}
\end{align}
which are sum rules valid upto first order in the small quark mixing angles. On
the other hand, performing the multiplication in the opposite order,
\ie,
\begin{equation}
W_2 = V_{\rm CKM} U_{\rm MNS},
\label{eq:W_2}
\end{equation}
we find for the total mixing angles of the $W_2$ matrix
\begin{align}
\theta_{12} &\simeq \theta_{12}^{\rm MNS} + \left( c_{23}^{\rm MNS}
\theta_{12}^{\rm CKM} - s_{23}^{\rm MNS} \theta_{13}^{\rm CKM} \right)
\sec \theta_{13}^{\rm MNS}, \label{eq:th12_2} \\
\theta_{13} &\simeq \theta_{13}^{\rm MNS} + s_{23}^{\rm MNS}
\theta_{12}^{\rm CKM} + c_{23}^{\rm MNS} \theta_{13}^{\rm CKM}, \\
\theta_{23} &\simeq \theta_{23}^{\rm MNS} + \theta_{23}^{\rm CKM} + 
\left( s_{23}^{\rm MNS} \theta_{13}^{\rm CKM} - c_{23}^{\rm MNS}
\theta_{12}^{\rm CKM} \right) \tan \theta_{13}^{\rm MNS}. \label{eq:th23_2}
\end{align}
Multiplying the two mixing matrices in the order $V_{\rm CKM} U_{\rm
MNS}$ has also been discussed in
\Refs~\cite{Xing:2005ur,Lindner:2005pk}. Note that the mixing angles
for the $W_2$ matrix in \eqs~(\ref{eq:th12_2})-(\ref{eq:th23_2}) can
be obtained from the mixing angles for the $W_1$ matrix in
\eqs~(\ref{eq:th12_1})-(\ref{eq:th23_1}) by replacing the ``12''
indices with the ``23'' indices, and vice versa. Therefore, in the
first case, the total mixing angle $\theta_{12}$ is linearly dependent
on all quark mixing angles and the other total mixing angles
$\theta_{13}$ and $\theta_{23}$ are only linearly dependent on
$\theta_{13}^{\rm CKM}$ and $\theta_{23}^{\rm CKM}$, whereas in the
second case, the total mixing angle $\theta_{23}$ is linearly
dependent on all quark mixing angles and the other total mixing angles
$\theta_{12}$ and $\theta_{13}$ are only linearly dependent on
$\theta_{12}^{\rm CKM}$ and $\theta_{13}^{\rm CKM}$. Furthermore, if
the CP-violating phases are assumed to be non-zero, then the resulting
formulas for the total mixing angles will be much more complicated
expressions.

In general, without any specific parameterization of the mixing
matrices, but instead using the matrix elements of the mixing
matrices, \eqs~(\ref{eq:W_1}) and (\ref{eq:W_2}) can be written as
$\sum_{k=1}^3 (U_{\rm MNS})_{ik} (V_{\rm CKM})_{kj} = (W_1)_{ij}$ and
$\sum_{k=1}^3 (V_{\rm CKM})_{ik} (U_{\rm MNS})_{kj} = (W_2)_{ij}$,
respectively, where $i$ and $j$ are fixed ($i,j = 1,2,3$). Assuming
that the CKM mixing matrix is close to the $3 \times 3$ identity
matrix, \ie, $(V_{\rm CKM})_{ij} = \delta_{ij} + \epsilon_{ij}$, where
$\delta_{ij}$ is Kronecker's delta and $\epsilon_{ij}$'s are small,
which should correspond to the quark mixing angles being small, we
obtain the following relations $(U_{\rm MNS})_{ij} + \sum_{k=1}^3
(U_{\rm MNS})_{ik} \epsilon_{kj} = (W_1)_{ij}$ and $(U_{\rm MNS})_{ij}
+ \sum_{k=1}^3 (U_{\rm MNS})_{kj} \epsilon_{ik} = (W_2)_{ij}$. {}From
these relations, inserting a specific parameterization of the mixing
matrices, it is then possible to derive similar sum rules to those
obtained in \eqs~(\ref{eq:th12_1})-(\ref{eq:th23_1}) and
(\ref{eq:th12_2})-(\ref{eq:th23_2}) for this specific
parameterization.

Inserting the best-fit values including the ranges of the quark and
leptonic mixing angles into the formulas of the total mixing angles
\eqs~(\ref{eq:th12_1})-(\ref{eq:th23_1}), we obtain the following
values for the mixing angles
$$
\theta_{12} = 46.2^\circ \pm 5.4^\circ, \quad \theta_{13} =
1.5^\circ \pm 12.8^\circ, \quad \mbox{and} \quad \theta_{23} =
46.9^\circ \pm 10.9^\circ,
$$
which should be compared with the ``exact'' numerical values that
actually are exactly the same. This means that the expansion formulas
are very accurate. Observe that the error propagation is completely
dominated by the errors in the leptonic mixing angles and that the
contribution from the errors in the quark mixing angles is, in
principle, negligible. Similar, inserting the best-fit values
including the ranges into \eqs~(\ref{eq:th12_2})-(\ref{eq:th23_2}), we
find that
$$
\theta_{12} = 42.3^\circ \pm 6.8^\circ, \quad \theta_{13} =
9.3^\circ \pm 14.3^\circ, \quad \mbox{and} \quad \theta_{23} =
47.4^\circ \pm 12.7^\circ,
$$
which also should be compared with the ``exact'' numerical values that
are $\theta_{12} = 42.3^\circ \pm 8.7^\circ$, $\theta_{13} = 9.3^\circ
\pm 14.2^\circ$, and $\theta_{23} = 46.7^\circ \pm 12.4^\circ$. Again,
the agreement between the results of the expansion formulas and the
``exact'' numerical calculations is very good. However, the errors are
slightly larger than in the previous case, but again completely
dominated by the contribution from the errors in the leptonic mixing
angles.

It is interesting to note that in both cases a bilarge (\ie, two
mixing angles are close to maximal or maximal and one angle is small
or zero) mixing pattern arises, which means that $\theta_{12} \simeq
45^\circ$, $\theta_{13} \simeq 0$, and $\theta_{23} \simeq
45^\circ$. In addition, both cases are even consistent with a
bimaximal mixing pattern, where $\theta_{12} = 45^\circ$, $\theta_{13}
\simeq 0$, and $\theta_{23} = 45^\circ$.

Finally, there is actually a third possible way (maybe even more
natural) to combine the two mixing matrices, which would be to take a
linear combination of the two matrices $W_1$ and $W_2$, \ie, $W_3 = a W_1
+ b W_2$, where $a$ and $b$ are constants. However, this would lead to
a total mixing matrix, which would not be unitary. Therefore, this
possibility will not be discussed any further.

\section{Discussion of earlier results}
\label{sec:discussion}

The QLC relation indicates that there could be a quark-lepton symmetry
or even quark-lepton unification based on the Pati--Salam model
\cite{Pati:1973rp,Pati:1974yy} such as SU(5) [or SU(5) and SO(10)]
GUT. Recently, this relation has been generally investigated by
Minakata and Smirnov \cite{Minakata:2004xt}. If not (numerically)
accidental, then a solid motivation for the QLC relation needs to be
found and it has to be rigorously experimentally tested. Furthermore,
renormalization group equations for running of the QLC relation have
been derived and analyzed in
\Refs~\cite{Antusch:2005gp,Lindner:2005pk}. The result of the analysis
suggest that if the QLC relation is assumed at high energies, then it
does not necessarily mean that the QLC relation is fulfilled at low
energies. Note that the leptonic mixing runs faster than the quark
mixing due to the fact that the leptonic mixing angles are larger than
the quark mixing angles.

In addition, Raidal \cite{Raidal:2004iw} has suggested three relations
$\theta_{12}^{\rm CKM} + \theta_{12}^{\rm MNS} = \tfrac{\pi}{4}$,
$\theta_{13}^{\rm CKM} \sim \theta_{13}^{\rm MNS} = {\cal
O}(\lambda^3)$, and $\theta_{23}^{\rm CKM} + \theta_{23}^{\rm MNS} =
\tfrac{\pi}{4}$ motivated by a flavor symmetry, which indicate that
there could exist a simple relation between the quark and leptonic
mixings. In principle, the relations proposed by Raidal serve as a
generalization of the QLC relation. In the derivation of these
relations, it has been assumed that the quark mixing matrix describes
the deviation of the leptonic mixing matrix from exactly bimaximal,
which he concludes should be due to some unknown underlying
non-Abelian flavor physics. Using the three relations, Li and Ma
\cite{Li:2005yj} have performed several test of these
relations. Especially, they have parameterized the MNS mixing matrix
with an assumed bimaximal mixing matrix as well as the Wolfenstein
parameters \cite{Wolfenstein:1983yz} of the CKM mixing matrix. Using
this parameterization, they have calculated both possible products of
the two mixing matrices and found theoretically that the relation
$U_{\rm MNS} V_{\rm CKM} = W_{\rm bimaximal}$ is in better agreement
with Raidal's relations than the relation $V_{\rm CKM} U_{\rm MNS} =
W_{\rm bimaximal}$, where $W_{\rm bimaximal}$ is the assumed bimaximal
mixing matrix. Note that the second relation $V_{\rm CKM} U_{\rm MNS}
= W_{\rm bimaximal}$ has also been discussed in
\Refs~\cite{Giunti:2002ye,Giunti:2002pp,Frampton:2004ud,Kang:2005as}.

However, in this paper, we have not assumed the exact relations of the
quark and leptonic mixing matrices used by Li and Ma, but we have instead
phenomenologically investigated the matrix products $U_{\rm MNS}
V_{\rm CKM}$ and $V_{\rm CKM} U_{\rm MNS}$, and hence, we have derived
series expansion formulas for the total mixing angles upto first order
in the small quark mixing angles. Using the present allowed
experimental and phenomenological ranges of the quark and leptonic
mixing angles, it has been found that bimaximal (or at least bilarge)
mixing naturally appears, \ie, we have not assumed that the product of
the mixing matrices is a bimaximal mixing matrix.

\section{Summary and conclusions}
\label{sec:S&C}

In summary, we have derived series expansion formulas for the
fermionic mixing angles in terms of the quark and leptonic mixing
angles. These formulas are explicit sum rules for the quark and
leptonic mixing angles in a unified fermionic picture of quark and
leptonic mixing. The formulas are valid upto first order in the small
quark mixing angles. However, due to the smallness of the quark mixing
angles, the formulas are indeed very accurate. In addition, we have
shown that it turns out, using these sum rules, that present data
naturally lead to bilarge or bimaximal fermion mixing. It is important
to note in this paper that we have not assumed bilarge or bimaximal
fermionic mixing, but bilarge or bimaximal fermionic mixing is purely
a result of combining the mixings stemming from experimental and
phenomenological data of quarks and leptons. The way to test the
results presented in this paper will be to use data from future
precision measurements of the quark and leptonic mixing angles such as
data from B physics and neutrino oscillation experiments.

\section*{Acknowledgments}

I would like to thank Mattias Blennow, Tomas H{\"a}llgren, and Gunnar
Sigurdsson for useful discussions. This work was supported by the
Royal Swedish Academy of Sciences (KVA), Swedish Research Council
(Vetenskapsr{\aa}det), Contract No.~621-2001-1611, 621-2002-3577, and
the G{\"o}ran Gustafsson Foundation (G{\"o}ran Gustafssons Stiftelse).

\end{document}